# Ultralow complexity long short-term memory network for fiber nonlinearity mitigation in coherent optical communication systems

Hao Ming[1,†], Xinyu Chen[1,†], Xiansong Fang[1,†], Lei Zhang[1], Chenjia Li[1], Fan Zhang[1, 2,*]

**Abstract**

Fiber Kerr nonlinearity is a fundamental limitation to the achievable capacity of long-distance optical fiber communication. Digital back-propagation (DBP) is a primary methodology to mitigate both linear and nonlinear impairments by solving the inverse-propagating nonlinear Schrödinger equation (NLSE), which requires detailed link information. Recently, the paradigms based on neural network (NN) were proposed to mitigate nonlinear transmission impairments in optical communication systems. However, almost all neural network-based equalization schemes yield high computation complexity, which prevents the practical implementation in commercial transmission systems. In this paper, we propose a center-oriented long short-term memory network (Co-LSTM) incorporating a simplified mode with a recycling mechanism in the equalization operation, which can mitigate fiber nonlinearity in coherent optical communication systems with ultralow complexity. To validate the proposed methodology, we carry out an experiment of ten-channel wavelength division multiplexing (WDM) transmission with 64 Gbaud polarization-division-multiplexed 16-ary quadrature amplitude modulation (16-QAM) signals. Co-LSTM and DBP achieve a comparable performance of nonlinear mitigation. However, the complexity of Co-LSTM with a simplified mode is almost independent of the transmission distance, which is much lower than that of the DBP. The proposed Co-LSTM methodology presents an attractive approach for low complexity nonlinearity mitigation with neural networks.

Fiber Kerr nonlinearity imposes a fundamental limitation to the achievable transmission distance and information capacity of optical fiber communication[1]. This limitation is mainly attributed to deterministic Kerr fiber nonlinearities and the interaction of fiber nonlinearity with amplified spontaneous emission noise from cascaded optical amplifiers[2-3], which is identified as a nonlinear Shannon capacity limit. Overcoming fiber nonlinearity is one of the most challenging tasks to extend the capacity and the transmission distance of optical fiber communication systems.

To mitigate nonlinear effects in optical fibers, several nonlinear compensation (NLC) techniques have been developed, such as digital back-propagation (DBP)[4], Volterra series filtering[5], perturbation-based compensation[6], optical phase conjugation (OPC)[7-8], and phase-conjugated twin waves (PCTW)[9]. DBP is a primary methodology to mitigate both linear and nonlinear impairments by solving the inverse-propagating nonlinear Schrödinger equation (NLSE). However, DBP is not feasible for commercial implementation due to its high complexity, especially for long-haul and/or multi-channel systems. The Volterra filtering and the perturbation method also demand high computation resources due to the operation of the summarization of nonlinear triplets. The requirement of additional optical apparatus for phase conjugation at the mid-span of the fiber link increases both the cost and the flexibility of the OPC technique. For PCTW, it is also impractical since the data rate is reduced by half due to the conjugated twin waves.

Recently, machine learning (ML) techniques have been proposed to mitigate nonlinear transmission impairments in optical communication systems. The learned DBP technique was proposed by treating multi split-step Fourier iterations as neural network (NN) layers for parameter optimization[10], which shows better performance than the conventional DBP method. NN with the designed nonlinear perturbation triplets was proposed to pre-distort symbols at the transmitter[11]. This approach has better performance than the filtered-DBP while achieving a slight complexity advantage. A complex NN approach designed with reference to the averaged Manakov equations was proposed to mitigate fiber nonlinearity[12]. The simulation results show that its performance is better than that of the DBP with 2 steps per span (StPs), but its complexity is much higher than that of DBP-2-StPs[12].

In addition, a bi-directional long short-term memory network (Bi-LSTM) was numerically studied for the mitigation of fiber nonlinearity in the coherent system[13] and the nonlinear Fourier transform (NFT) based system[14]. The main advantage of Bi-LSTM is that they can efficiently handle inter-symbol-interference (ISI) among preceding and succeeding symbols caused by chromatic dispersion. However, Bi-LSTM demands high computation resources due to its iterative calculation based on time steps. Thousands or even more multiplications are required in long-distance optical transmission systems with Bi-LSTM equalization.

[1]*State Key Laboratory of Advanced Optical Communication Systems and Networks, Frontiers Science Center for Nano-optoelectronics, Department of Electronics, Peking University, Beijing 100871, China*
[2]*Peng Cheng Laboratory, Shenzhen 518055, China*
*Corresponding author: fzhang@pku.edu.cn
†These three authors contributed equally to this work.

In this paper, we propose a novel center-oriented long short-term memory network (Co-LSTM) scheme incorporating a simplified mode based on a recycling mechanism for ultralow complexity operation. We verify our method in an experiment of 10×512Gb/s wavelength division multiplexing (WDM) transmission over 1600km standard single-mode fiber (SSMF) with polarization division multiplexing (PDM) 16-ary quadrature amplitude modulation (16-QAM). The experiments show that the Co-LSTM can effectively mitigate fiber nonlinearity with a 0.51dB $Q^2$ factor gain and reduce the complexity to 5.2% of that of the Bi-LSTM and 28.4% of that of the DBP with 1 step per span. Our results suggest emerging low-complexity neural network-based nonlinear equalization in coherent optical transmission systems.

**The principle of LSTM.** Recurrent neural networks (RNNs) are mainly used to process sequence data, which predict the output of the current state by encoding the current and the previous data. However, standard RNNs face a major practical problem: the gradient of the total output error with respect to previous inputs quickly vanishes as the time delays between relevant inputs and errors increase[15]. LSTM improves the internal structure of the standard RNNs and solves the problem of vanishing gradients[16]. Moreover, LSTM can lean long-term dependencies by enforcing constant error flow through constant error carousels (CECs) within special units[16,17].

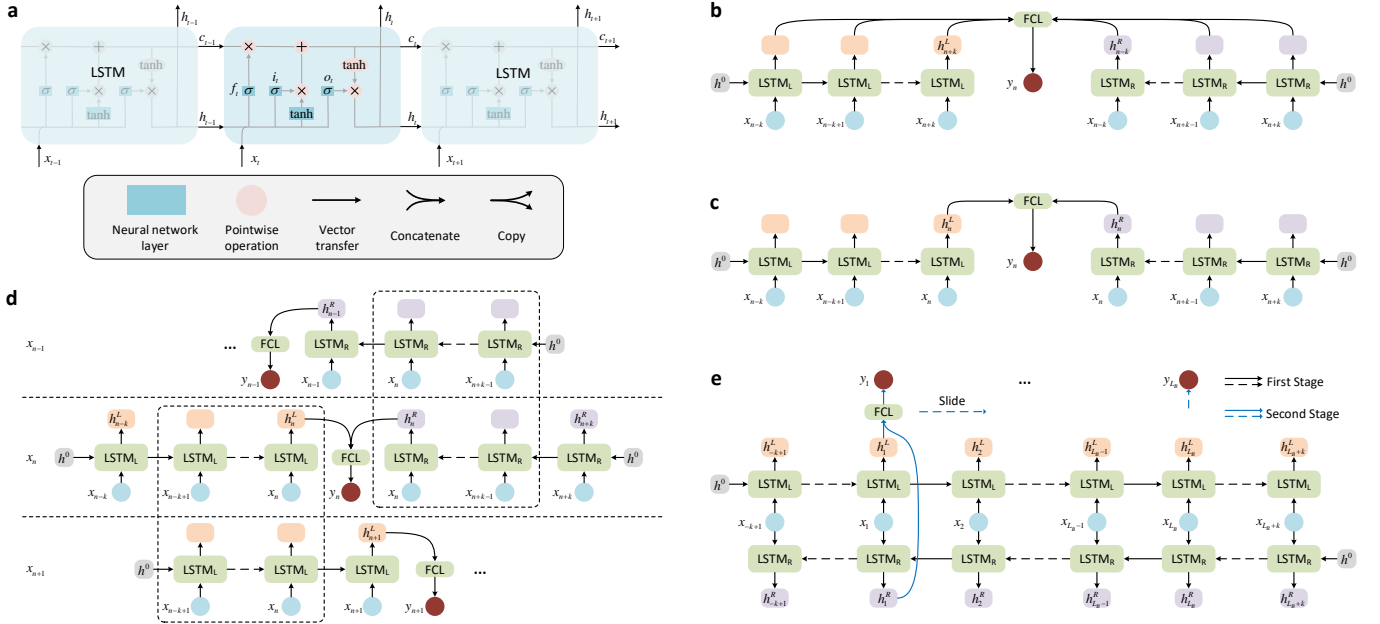

Fig. 1. (a) The structure (up) and the description (down) of LSTM. (b) The conventional Bi-LSTM equalization scheme and (c) the proposed Co-LSTM equalization scheme. (d) The calculation process of the Co-LSTM (standard mode). (e) The simplified calculation mode of the Co-LSTM. $x$ is the received symbol; $x_n$ is the target symbol; $y_n$ is the output symbol, $h^0$ is the zero vector.

The LSTM structure is shown in Fig. 1 (a). At the moment of $t$, LSTM is composed of an input vector $x_t$, cell state $c_t$, temporary state $c_t$, hidden state $h_t$, forget gate $f_t$, input gate $i_t$, and output gate $o_t$. The main idea of LSTM is that the cell state $c_t$ containing past information does not need to be updated at every moment, which is achieved by forget gate and input gate. The forget gate determines which past information in $c_{t-1}$ needs to be forgotten, and the input gate determines which current information in $c_t$ needs to be added. At last, the output gate determines which information in $c_t$ needs to be output.

The internal calculation process of the LSTM is as follows.

$$f_t = \sigma\left(\mathbf{w}_f\left[h_{t-1}, x_t\right] + b_f\right) \quad (1)$$

$$i_t = \sigma\left(\mathbf{w}_i\left[h_{t-1}, x_t\right] + b_i\right) \quad (2)$$

$$c_t = \tanh\left(\mathbf{w}_c\left[h_{t-1}, x_t\right] + b_c\right) \quad (3)$$

$$c_t = f_t \odot c_{t-1} + i_t \odot c_t \quad (4)$$

$$o_t = \sigma\left(\mathbf{w}_o\left[h_{t-1}, x_t\right] + b_o\right) \quad (5)$$

$$h_t = o_t \odot \tanh(c_t) \tag{6}$$

It is worth noting that the parameters of LSTM are independent of the time step $t$.

**The design of center-oriented LSTM equalization scheme.** Design an LSTM equalization scheme with reduced complexity while ensuring good performance is a challenging task. There are two ways to reduce the complexity of the LSTM equalization scheme. The first is to use a neural network with a simpler internal structure, such as a gated recurrent unit (GRU)[18]. The other is to simplify the external calculation structure of LSTM. This paper focuses on the second way. So far, most of the current works directly use a conventional Bi-LSTM scheme[13,14,19], which demands high computation resources. In this paper, we propose a center-oriented LSTM (Co-LSTM) scheme for simplifying the external calculation structure of Bi-LSTM. Both Bi-LSTM and Co-LSTM schemes are shown in Fig. 1 (b) and Fig. 1(c).

The conventional Bi-LSTM is composed of two unidirectional LSTMs both for the input symbols $x_{n-k}$ to $x_{n+k}$. $k$ is the single-side tap length determined by the ISI in a specific system. The hidden state of each time step is concatenated and then mapped out through a fully connected layer (FCL). In contrast, Co-LSTM consists of two unidirectional LSTMs respectively for the input symbols $x_{n-k}$ to $x_n$ (LSTM$_L$) and $x_n$ to $x_{n+k}$ (LSTM$_R$). LSTM$_L$ and LSTM$_R$ separately process the symbol information on the left and the right sides of the target symbol $x_n$. Finally, the hidden states of the last time step on the left and the right are concatenated and then mapped out through an FCL.

**Calculation Mode of Co-LSTM.** Fig. 1(d) shows the calculation process when using a Co-LSTM to equalize the symbols $x_{n-1}$, $x_n$, and $x_{n+1}$. To explain the calculation process concisely, the cell state is omitted in Fig. 1(d). When equalizing the target symbol $x_n$, the left side hidden state $h_n^L$ is calculated with the following equation.

$$\begin{aligned} h_n^L \big|_{y_n} &= \text{LSTM}_L \left( h^0, x_{n-k}, k+1 \right) \\ &= \text{LSTM}_L \left( h_{n-k}^L \big|_{y_n}, x_{n-k+1}, k \right) \end{aligned} \tag{7}$$

Here $h_n^L \big|_{y_n}$ is the $h_n^L$ that appears during the calculation of the output $y_n$. Eq. (7) indicates that using $h^0$ as the initialization state, LSTM$_L$ iteratively calculates $k+1$ steps forward from the symbol $x_{n-k}$. Similarly, when the symbol $x_{n+1}$ is equalized, $h_{n+1}^L$ can be calculated as

$$\begin{aligned} h_{n+1}^L \big|_{y_{n+1}} &= \text{LSTM}_L \left( h^0, x_{n-k+1}, k+1 \right) \\ &= \text{LSTM}_L \left( \text{LSTM}_L \left( h^0, x_{n-k+1}, k \right), x_{n+1}, 1 \right) \\ &= \text{LSTM}_L \left( h_n^L \big|_{y_{n+1}}, x_{n+1}, 1 \right) \end{aligned} \tag{8}$$

It can be seen that there are similar terms in Eq. (7) and Eq. (8). Those terms are marked by the left dash box in Fig. 1(d). We can prove that the following approximate relationship holds (see Supplement material for details).

$$\begin{aligned} h_n^L \big|_{y_n} &= \text{LSTM}_L \left( h_{n-k}^L \big|_{y_n}, x_{n-k+1}, k \right) \\ &\approx \text{LSTM}_L \left( h^0, x_{n-k+1}, k \right) \\ &= h_n^L \big|_{y_{n+1}} \end{aligned} \tag{9}$$

Then Eq. (8) can be rewritten as
$$h_{n+1}^L \big|_{y_{n+1}} \approx \text{LSTM}_L \left( h_n^L \big|_{y_n}, x_{n+1}, 1 \right). \tag{10}$$

In Eq. (10), $h_n^L \big|_{y_n}$ is used to calculate $h_{n+1}^L \big|_{y_{n+1}}$, which is the so-called recycling mechanism in our methodology. In doing so, the calculation of $h_{n+1}^L \big|_{y_{n+1}}$ is reduced from $k+1$ iterations required by Eq. (8) to only one iteration required by Eq. (10). Consequently, the computational complexity will be greatly reduced. The calculation process of $h_{n-1}^R \big|_{y_{n-1}}$ is similar to that of $h_{n+1}^L \big|_{y_{n+1}}$.

Based on the above-mentioned recycling mechanism, the calculation process of the standard mode that described in Fig. 1(d) can be simplified to the so-called simplified mode that described in Fig. 1(e).

As shown in Fig. 1(e) the simplified mode consists of two stages. In the first stage, for $L_B$ received symbols, we first use LSTM$_L$ and LSTM$_R$ to calculate the hidden states $h^L$ and $h^R$ at each moment (the black solid and the black dash lines). In the second stage, the hidden states $h^L$ and $h^R$ at the same time are concatenated and then mapped out through an FCL to obtain the equalized output $y$ (the blue solid and the blue dash lines).

For implementation, the Co-LSTM are trained and then used to equalize the received signals. The Co-LSTM adopts the standard mode in the training phase and the simplified mode in the equalization phase. Consider a data block with $L_B$ symbols, in the standard mode, the gradient only needs to be calculated $k$ steps back through the backpropagation through time (BPTT) algorithm[20], while in the simplified mode, $L_B - 1$ steps back need to be calculated (usually $L_B \gg k$). The simplified mode will greatly increase the computational complexity of the gradient and the storage required during the training phase. Therefore, we only use the simplified mode in the equalization phase that does not involve gradient calculation. It should be noted that an additional time delay will be introduced when the simplified mode is used for equalization. This is because, in the simplified mode, LSTM$_R$ needs to be calculated back from the end of the $L_B$ received symbols. Therefore, a data sequence can be divided into multiple blocks with an optimized length of $L_B$ to handle the time delay issue when performing the proposed Co-LSTM equalization scheme.

**Co-LSTM for WDM transmissions.** To validate the proposed Co-LSTM equalization method, we carry out an experiment of high-speed WDM transmission systems with 10 × 64 Gbaud PDM 16-QAM signals with 70 GHz channel spacing. The transmitted optical spectra of the signal after different distances are shown in Fig. 2 (a) with a resolution of 0.02 nm. The sixth WDM channel is chosen as the target one for evaluating the system performance in the following experiments.

Table 1 shows the complexity and the performance of Bi-LSTM and Co-LSTM. We use the number of real multiplications per bit (RMPB) to describe the complexity of the equalization scheme. $N$ is the order of the modulation format. $L_T$ is the tap length ($L_T = 2k + 1$). $C_L$ is the number of real multiplications of LSTM in a single time step, which can be expressed as follows.

$$C_L = 4n_{hidden}\left(n_{input} + n_{hidden}\right) + 3n_{hidden} \quad (11)$$

Here $n_{input}$ and $n_{hidden}$ are the number nodes of the input and the hidden layers of the LSTM, respectively. The measured $Q^2$ factors of the target channel are also shown for both two LSTM schemes after 1600 km SSMF transmission. The launch power is 1.0 dBm per channel. The relationship between the $Q^2$ factor and the measured bit-error-rate (BER) can be described by $Q^2 = 20\log_{10}[\sqrt{2}erfc^{-1}(2BER)]$. From Table 1 we can conclude that Co-LSTM with the standard mode reduces the complexity by nearly half while ensuring that the performance is slightly better than that of Bi-LSTM.

To further reduce the complexity of Co-LSTM equalization, we can apply the simplified mode. It can be seen that the performance of Co-LSTM with the simplified mode is similar to that of Co-LSTM with the standard mode, which verifies the correctness of Eq. (9). If the equalization phase of Co-LSTM adopts the standard mode, its complexity roughly linearly increases with the tap length. In contrast, when Co-LSTM adopts the simplified mode, its complexity will approach a constant as $L_B$ increases. In the following texts and figures, the equalization phase of Co-LSTM adopts the simplified mode.

The relationships between the complexity and the tap lengths of Bi-LSTM equalization and Co-LSTM equalization with the simplified mode are shown in Fig. 2 (b). We set $L_B = 30000$, $n_{input} = 4$, and $n_{hidden} = 16$. The complexity of Bi-LSTM linearly increases with the tap length, while the Co-LSTM complexity is almost independent of the tap length. Consequently, the complexity ratio $R_{Co/Bi}$ of Co-LSTM to Bi-LSTM is approximately $1/L_T$.

**Table 1. The complexity and the performance of different equalization schemes.**

| Equalization scheme | Training phase | Equalization phase | RMPB | $Q^2$ factor/dB |
|---|---|---|---|---|
| w/o LSTM | / | / | / | 6.48 |
| Bi-LSTM | / | / | $\frac{1}{\log_2 N}(2L_T C_L + 4L_T n_{hidden})$ | 6.94 |
| Co-LSTM | Standard mode | Standard mode | $\frac{1}{\log_2 N}((L_T + 1)C_L + 4n_{hidden})$ | 7.02 |
| Co-LSTM | Standard mode | Simplified mode | $\frac{1}{\log_2 N}\left(\frac{2L_B}{L_B - L_T + 1}C_L + 4n_{hidden}\right)$ | 6.99 |

Fig. 2 (c) shows the $Q^2$ factor of the target channel versus the launch power per channel after 1600 km SSMF transmission. With nonlinearity mitigation, the optimal launch power is 1.0 dBm per channel, which is 1.0 dB larger compared with that of the chromatic dispersion compensation (CDC). We perform DBP equalization with 1 step and 3 steps per span (DBP-1-StPs and DBP-3-StPs). The further gain of DBP with more steps per span is negligible in our experiments. Compared to the CDC only, DBP-1-StPs, DBP-3-StPs, and Co-LSTM have 0.42 dB, 0.55 dB, and 0.51 dB $Q^2$ factor improvement, respectively, at the optimal launch power of 1.0 dBm per channel. In addition, the typical constellations with different equalization schemes at 1.0 dBm per channel are shown in Fig. 2 (d).

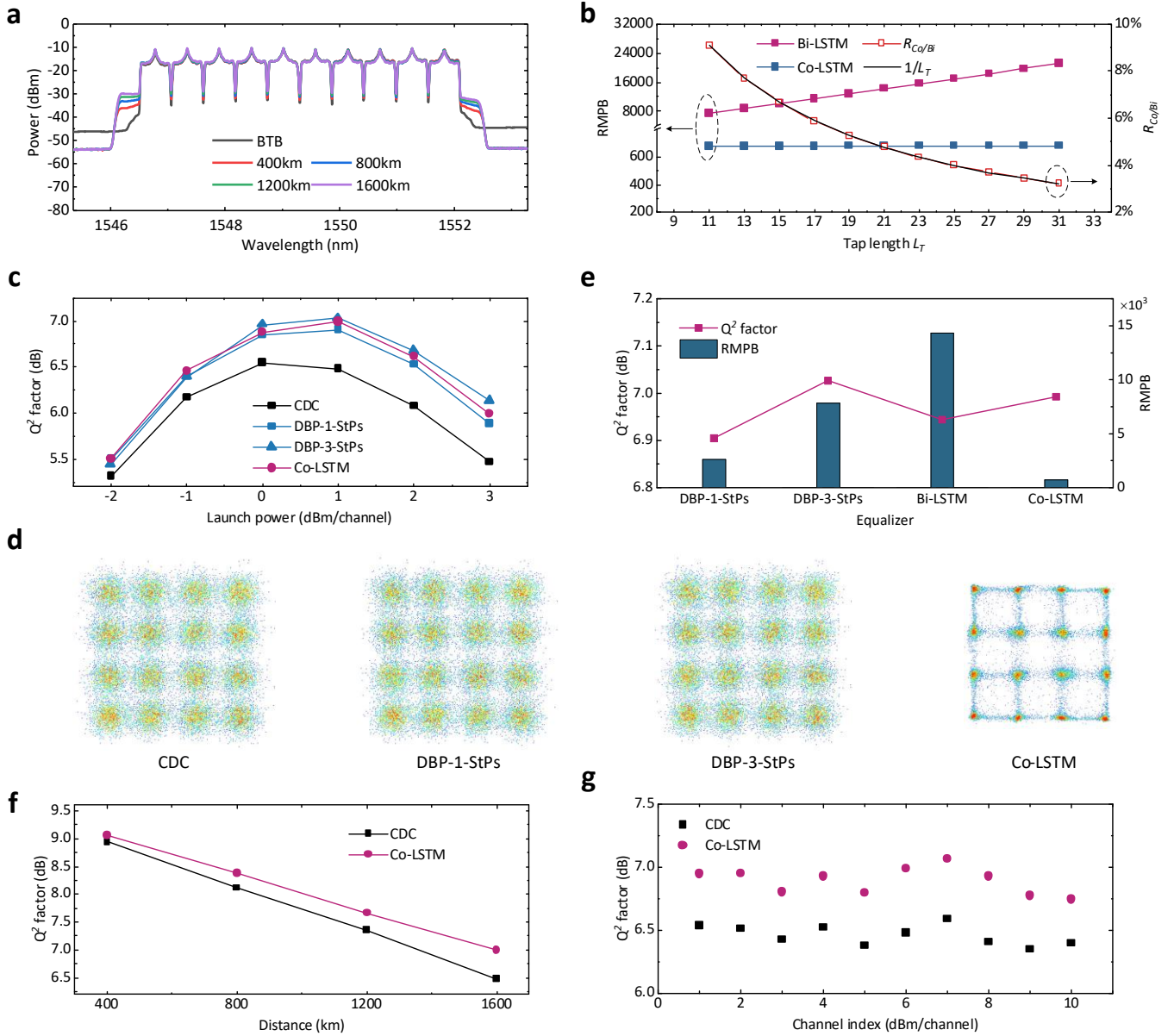

Fig. 2. (a) Spectra of the Nyquist-WDM signal in the case of a back-to-back, 400 km, 800 km, 1200 km, and 1600 km SSMF transmission. (b) The relationship between the complexity and the tap length of Bi-LSTM equalization and Co-LSTM equalization with the simplified mode. (c) $Q^2$ factor of the target channel versus launch power per channel. (d) The typical constellations with different equalization schemes at the launch power of 1.0 dBm per channel. (e) The performance and the complexity of different equalization schemes at the launch power of 1.0 dBm per channel. (f) $Q^2$ factor of the target channel versus the transmission distance with the launch power of 1.0 dBm per channel. (g) $Q^2$ factor for the ten WDM channels after 1600 km SSMF transmission.

The performance and the complexity of different equalization schemes are compared in Fig. 2 (e). The complexity of DBP can be expressed as follows.

$$C_{DBP} = \frac{4}{\log_2 N} n_{span} n_{step} n_{up} \left(2\left(\log_2 n_{FFT} + 1\right) + 1\right) \qquad (12)$$

Here $n_{span}$ is the number of fiber spans. $n_{step}$ is the number of steps per span used. $n_{up}$ is the oversampling ratio. $n_{FFT}$ is the FFT size. For a fair comparison, we also take the complexity of CDC into account when calculating the complexity of LSTM based equalization. We chose an optimized value of $L_T = 21$. It can be seen from Fig. 2 (e) that the performance of nonlinearity mitigation of Co-LSTM is between those of DBP-1-StPs and DBP-3-StPs. In addition, Co-LSTM reduces the complexity to 5.2% of that of Bi-LSTM (28.4% of that of the DBP-1-StPs) while achieving slightly better performance than Bi-LSTM. It should be noted that the DBP complexity increases with the transmission distance, while the complexity of Co-LSTM equalization with the simplified mode is almost independent of the fiber link length. For longer fiber link, the complexity ratio of Co-LSTM to DBP will be further reduced.

Fig. 2 (f) shows the $Q^2$ factor of the target channel versus the transmission distance at the optimal launch power of 1.0 dBm per channel. When the distance increases from 400 km to 1600 km, Co-LSTM offers 0.12 dB to 0.51 dB Q2 factor improvement compared to CDC. Fig. 2 (g) shows the $Q^2$ factor of all the ten WDM channels after 1600 km SSMF transmission at the optimal launch power of 1.0 dBm per channel. The average $Q^2$ factors are 6.46 dB and 6.89 dB, respectively, for the signal with the CDC only and Co-LSTM equalization with the simplified mode.

**Discussion**

In this paper, we propose a novel LSTM scheme that is so-called Co-LSTM, which can incorporate a simplified mode into the equalization phase. With the help of Co-LSTM, 10 × 512Gb/s PDM 16-QAM Nyquist-WDM signals are transmitted over 1600 km SSMF. The experiment results show that Co-LSTM can mitigate fiber nonlinearity efficiently and offer a 0.51 dB $Q^2$ factor improvement compared with the CDC only, which is comparable to the performance of DBP. For the specific system in our experiment, the complexity of Co-LSTM is only 5.2% of that of conventional Bi-LSTM and 28.4% of that of DBP-1-StPs. To our best knowledge, this is the first time that a neural network-based equalization has been implemented with excellent performance while its complexity is significantly lower than that of the DBP method. It should be noted that, based on a recycling mechanism in the equalization phase, the proposed Co-LSTM with a simplified mode has a computation complexity almost independent of the transmission distance, which shows an intrinsic advantage over the DBP method. We believe that the concept of Co-LSTM together with the simplified methodology should provide general guidance for the development of low-complexity neural network-based equalization techniques.

**Methods**

**Experimental setup.** The experimental setup is shown in Fig. 3. At the transmitter side, the loading channels are from 10 external cavity lasers (ECLs, EXFO IQS-2800) at frequencies from 193.16 THz to 193.79 THz, which are then switched off in turn before multiplexing with the channel under test (CUT) using a Waveshaper (WS, Finisar Waveshaper 1000s). The Waveshaper helps to equalize the optical power of each channel during multiplexing. The linewidth of each laser is ~100 kHz. The baseband Nyquist 64GBaud 16QAM signal is generated by an arbitrary waveform generator (AWG, Keysight M8194A, with a 3dB bandwidth of ~45 GHz) operating at 120 GSa/s. After amplified by a pair of electrical amplifiers (EAs, SHF S807) with 50GHz bandwidth, the waveforms from AWG are used to drive the IQ modulators (IQ Mod.1/2 with a 3dB bandwidth of ~27 GHz) for the CUT and other WDM channels, respectively. PDM is emulated with a polarization beam splitter/combiner (PBS/PBC) and an optical delay line. Before the signals are launched into the recirculating fiber loop, an erbium-doped fiber amplifier (EDFA) is applied to adjust the launch power.

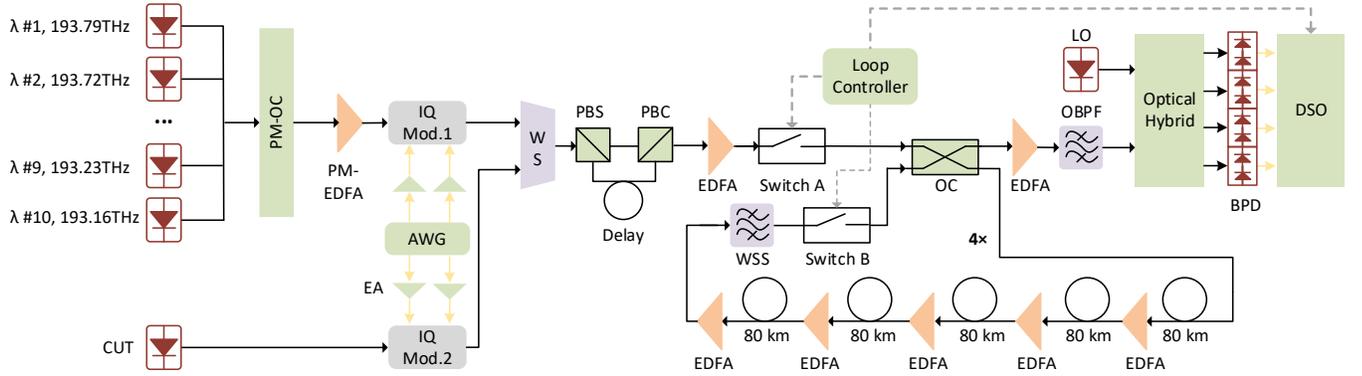

Fig. 3. Experimental setup. ECL: external cavity laser; PM-OC: polarization-maintaining optical coupler; AWG: arbitrary waveform generator; EA: electrical amplifier; IQ-Mod: IQ-modulator; WS: waveshaper; EDFA: erbium-doped fiber amplifier; SSMF: standard single-mode fiber; OBPF: optical band-pass filter; LO: local oscillator; DSO: digital storage oscilloscope.

The fiber loop has a length of 400 km, which consists of five spans of SSMF with EDFA-only amplification. A wavelength selective switch (WSS, Finisar Wavelength Selective Switch) is employed to suppress amplified spontaneous emission (ASE) noise accumulation out of the WDM band. At the receiver side, the signal is first amplified by an EDFA to adjust the received power. Then an optical band-pass filter (OBPF, Yenista Optics XTM-50) is utilized to select the desired channel. After that, the signal and the local oscillator (LO) are sent to an optical 90° hybrid followed by balanced photodiodes (BPDs, Finisar BPDV3120R) with ~70 GHz 3-dB bandwidth. Finally, the received signals are amplified with EAs (SHF S807) and then sampled by an 80 GSa/s real-time oscilloscope (Keysight DSA-X 96204Q) to perform off-line digital signal processing (DSP).

**Digital signal processing flows.** The DSP flows are shown in Fig. 4 (a) and Fig. 4 (b). In the Tx-side DSP, the bit stream is mapped to 16-QAM first. After 15 times up-sampling, the signal is digitally shaped using a root raise cosine (RRC) filter with a roll-off factor of 0.01. Then the signal is 8 times down-sampled. Next, the Quadratic Curve (QC) fitting method[21] is employed in the digital pre-emphasis process to compensate for the linear response of the digital-analog-converters (DACs) and electrical drivers.

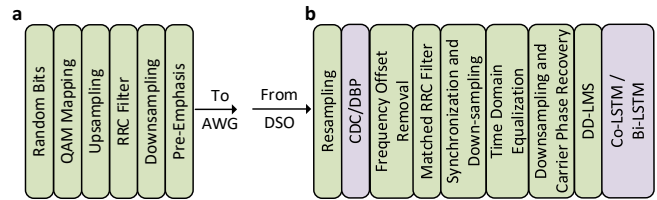

Fig. 4. Structure of (a) Tx-side DSP and (b) Rx-side DSP.

For the Rx-side DSP, the signal is firstly re-sampled to 4 samples per symbol (SPS). Then the chromatic dispersion compensation or the DBP is applied. The frequency offset between the LO and the signal is estimated by obtaining the maximum of $|FFT[r^4(t)]|$. Here $FFT(\cdot)$ is the fast Fourier transformation and $r(t)$ is the signal after down-conversion. After the matched filter, synchronization, and down-sampling to 2 SPS, time-domain equalization (TDE) is performed with a finite impulse response (FIR) filter. The filter taps are calculated from the training sequences with the recursive least square (RLS) algorithm. After down-sampling to 1 SPS, the carrier phase recovery is conducted in two steps: 1) Coarse phase tracking is realized by using the uniformly distributed pilot symbols; 2) Accurate phase recovery is based on the blind phase search (BPS) algorithm[22]. Then a decision-direct LMS (DD-LMS) filter is used to improve the signal quality by tracing the time-varying channel response. After that, Co-LSTM or Bi-LSTM can be adopted for nonlinear mitigation. Co-LSTM and Bi-LSTM use the same training and test sets, with lengths of $2\times10^4$ and $2\times10^5$ symbols, respectively. Finally, the BER is calculated by error counting based on the test sets.